\documentclass{aa}

\usepackage[authoryear]{natbib}
\bibpunct{(}{)}{,}{a}{}{;} 

\usepackage{graphicx}
\newcommand{\msun}{{M$_\odot$}}
\newcommand{\lsun}{L$_\odot$}

\newcommand{\kms}{km\,s$^{-1}$}
\newcommand{\halpha}{H$\alpha$}
\newcommand{\cm}{cm$^{-2}$}
\newcommand{\hii}{H{\sc ii}}

\begin{document}
 
\title{Molecular and
  ionized gas in the tidal tail in Stephan's Quintet}

\titlerunning{Molecular gas and ionized gas the tidal tail in 
Stephan's Quintet}

\author{
Ute Lisenfeld\inst{1,2}
\and Jonathan~Braine\inst{3}
\and Pierre-Alain Duc\inst{4}
\and Elias Brinks\inst{5}
\and Vassilis Charmandaris\inst{6,7}
\and St\'ephane Leon\inst{1}
} 

\institute{Instituto de Astrof\'\i sica de Andaluc\'\i a, CSIC, Apdo. 3004,
 18040 Granada, Spain 
\and Dept. de F\'\i sica Te\'orica y del Cosmos, 
Universidad de Granada, Granada, Spain
\and Observatoire de Bordeaux, UMR 5804, CNRS/INSU, B.P. 89, 
  F-33270 Floirac, France
\and CNRS URA 2052 and CEA/DSM/DAPNIA Service d'Astrophysique, Saclay,
91191 Gif sur Yvette Cedex, France
\and INAOE, Apdo. Postal 51 \& 216, Puebla, Pue 72000, Mexico
\and Cornell University, Astronomy Department, Ithaca, NY 14853, USA
\and Chercheur Associ\'e, Observatoire de Paris, LERMA, 61 Av. de l'Observatoire, 75014 Paris, 
France
}

\date{Received   / Accepted   }

\authorrunning{Lisenfeld et al.}
 
\abstract{We have mapped with the IRAM interferometer at Plateau de Bure 
(PdBI) the $^{12}$CO emission towards intergalactic star forming regions 
located in the tidal tail stemming from NGC 7319, in the Stephan's Quintet 
compact group of galaxies. The  $^{13}$CO emission of the same region was 
observed with the IRAM 30m telescope and optical spectroscopy
of several \hii \ regions in the area were obtained with the Calar Alto 
3.5m telescope.We recovered with the interferometer about 50\% of the  
$^{12}$CO(1--0) total emission that had been earlier measured with single 
dish observations \citep{Lisenfeld02}, indicating that about half of the 
molecular gas is distributed on spatial scales larger than about 10-15"
(corresponding to 4-6 kpc) to which PdBI is not sensitive. We find two main 
areas of CO emission: (i) an elongated region towards the area
known as SQ~B where a Tidal Dwarf Galaxy could currently be forming
(ii) a barely resolved area at the tip of the optical tidal arm. Both regions 
follow dust lanes visible on HST images and their 
CO peak coincides spatially exactly with the maximum of the \halpha \ line 
emission. In SQ~B, there is furthermore very good
kinematical agreement between the CO, \halpha \ and HI components.
We conclude from these coincidences that the gaseous matter found in 
quantities in the area { is}
physically associated to the optical tidal tail 
and thus that the intergalactic atomic hydrogen there was expelled from 
NGC~7319. Its origin had previously been much debated. 
Furthermore, the relatively high oxygen abundances (about solar)
estimated from the optical spectra of the \hii \ regions imply that the gas
feeding the star formation originated from the inner regions of the parent 
galaxy. In SQ~B, we derive from different tracers a star formation rate, 
corrected for dust extinction - which is important in the area - of 
0.5 \msun/yr, i.e. one of the highest values so far measured outside galaxies.
The inferred molecular gas consumption time of 0.5 Gyr
lies in the range of values found for spiral and starburst galaxies. 
On the other hand, the ratio of  $^{12}$CO/$^{13}$CO $> 25$ is much higher
than the values found in disks of spiral galaxies. 
A relatively low opacity  for the $^{12}$CO gas is the most likely reason.
}
 
\maketitle
 
\section{Introduction}

The  
Hickson Compact Group Stephan's Quintet (Hickson Compact Group 92;
hereafter SQ)  consists of 
four interacting  galaxies (NGC~7319,
NGC~7318a, NGC~7318b, and NGC~7317) and a foreground galaxy (NGC 7320).
A fifth galaxy, NGC 7320c, situated $\sim$ 
4 arcmin to the east, is also dynamically 
associated to the group. 
The group has
experienced a violent dynamical history with numerous interactions
between the different members during the past Gyr (see \citet{Moles97}
and \citet{Sulentic01} for plausible scenarios).  As a result of these
interactions, two tidal arms, a faint older one, and a brighter young
one stemming from NGC~7319, have been created towards 
the eastern side of the group.  \citet{Sulentic01} suggest that
each of these tails was created by a passage of NGC~7320c, the galaxy
towards which both tails are pointing.

Several
knots of star formation (SF) are visible in the young tidal arm
extending from NGC~7319 to the east. The brightest region, identified
as B by \citet{Xu99} and hereafter called SQ~B, shows also
mid-infrared and H$\alpha$ emission. This region has been identified
as a good candidate for a Tidal Dwarf Galaxy (TDG)
\citep{Lisenfeld02}.  Another fainter SF region is visible at the very
tip of the tidal tail, hereafter called SQ~tip. 

One of the most striking properties of this group is that the major
part of the gas is in the intragroup medium. 
Abundant atomic gas is present
to the east of the three central galaxies.
\citet{Williams02} concluded from the HI kinematics 
that in reality this gas clouds consist of two subclouds, 
which they called Arc-N and Arc-S. Arc-N coincides at its 
southern end with the young tidal tail and Arc-S largely overlaps
with the old tidal tail. 
In the northern part of Arc-N several 
\hii \  regions are detected \citep{Sulentic01,Mendes04}.
The origin of the gas is unclear. 
A plausible explanation for the gas in Arc-N
is that it  has been stripped
from NGC~7319, the galaxy where the young tidal tail starts
\citep{Sulentic01}. A definite physical relation between Arc-N
and the young stellar tail is however not proven, the apparent overlap
could in principle be a projection effect.
The only argument in favor of a real association 
is the fact that the sharp inner edge of the gas
cloud follows the shape of  the stellar optical tail \citep{Sulentic01}.
Part of the HI could also have been stripped from NGC~7320c, the other
spiral galaxy in the group which completely lacks atomic gas.

 \citet{Lisenfeld02} observed the region around SQ~B, close to the
peak of the HI distribution, with the
IRAM 30m telescope and found abundant molecular gas ($7\times
10^8$\msun) covering a huge area of about 20 kpc.  
%
In this paper we present new millimeter observations of $^{12}$CO with the
IRAM interferometer at Plateau de Bure covering the region around
SQ~B and SQ~tip.  The goal of these
observations was to map the distribution of the molecular gas at a
higher spatial resolution and to study its relation to the SF regions
in the tidal tails. Furthermore, we obtained optical spectroscopy of
SQ~B and SQ~tip in order to further study the physics of the 
star forming regions, as well as
observations  of SQ~B with the IRAM 30m telescope of $^{13}$CO in order
to collect more information about the physical state of the molecular
gas.  

In Sect. 2, we describe the observations; in Sect. 3, we present the results;
in Sect. 4, we discuss some of their implications and
Sect. 5 gives a summary of our
most important results and conclusions.
We adopt a distance of 85\,Mpc, based on a recession velocity of
6400 \kms \ to SQ and H$_0$=75\,km\,s$^{-1}$\,Mpc$^{-1}$, in which
case 10\arcsec \ correspond to 4.1\,kpc.

\section{Observations and data reduction}  

\subsection{Plateau de Bure Interferometer}

Observations of the eastern 
young tidal tail of SQ were carried out with the IRAM 
interferometer at Plateau de 
Bure (PdBI)\footnote{Based on observations carried out
with the IRAM Plateau de Bure Interferometer. IRAM is supported by
INSU/CNRS (France), MPG (Germany) and IGN (Spain).}
 between May and December 2002,
using the CD set of configurations of the array.

We observed simultaneously the J=1--0 and J=2--1 lines of $^{12}$CO in
a single field centered at $\alpha _{\rm J2000}=22^{\rm h}36^{\rm
  m}11.1^{\rm s}$ and $\delta _{\rm J2000}=33^{\circ }57'17''$.  The
primary beam size which gives the diameter beyond which the
sensitivity of the instruments falls below half of its sensitivity
at the center,  is 44$''$ (22$''$) in the 1--0 (2--1) line.  The
spectral correlator was split in two halves centered at 112.788 and
225.572 GHz, respectively, i.e., the transition rest frequencies
corrected for an assumed recession velocity of $v_{\rm o}{\rm
  (LSR)}=6600$ \kms. We observed each line in parallel
with two correlator configurations:
a high-resolution
configuration with a frequency resolution of 0.312 MHz  
and a bandwidth of 80 MHz , corresponding to a velocity
resolution of 0.83 \kms \ and a bandwidth of 212 \kms \ at CO(1--0),
and a low-resolution configuration with a frequency resolution of 1.28 MHz  
and a bandwidth of 320 MHz , corresponding to a velocity
resolution of 3.32 \kms \ and a bandwidth of 830 \kms \ at CO(1--0).
Since the high-resolution data were rather noisy, we only use 
the low-resolution data cube in this paper.
Phase and amplitude calibrations were performed by observing the
nearby quasars 2145+065, 2201+315, 2234+282, 3C454.3, 3C273 and
1633+382, as well as MWC349.  The fluxes of the primary calibrators
were determined from IRAM 30m measurements and taken as an input to
derive the absolute flux density scale in our map; the latter is
estimated to be accurate to 10\%.
We estimate the uncertainty in the final flux calibration to be
about 20\%.


The data of the $^{12}$CO(2--1) line were of poor quality and we will
not use them in the analysis presented in this paper.  The image
reconstruction was done using standard IRAM/Gildas software.  
We used natural weighting and no taper to generate the 
$^{12}$CO(1--0) line maps with a 0.5\arcsec \ sampling.
 The
corresponding synthesized beam is $4.33''\times 3.46''$, $ {\rm
  PA}=-85^{\circ }$.  During the process of data reduction we have
also tested uniform weighting and different taperings in order to
search for weaker emission features, but the resulting maps were not
significantly different.  The data, except for the displayed
maps, are  corrected for primary beam
attenuation.

\subsection{IRAM 30m telescope}

We observed the $^{13}$CO(1--0) and $^{13}$CO(2--1) towards SQ B
(centered at $\alpha _{\rm J2000}=22^{\rm h}36^{\rm m}10.3^{\rm s}$
and $\delta _{\rm J2000}=33^{\circ }57'17''$) with the IRAM 30m
telescope on Pico Veleta in March 2004 under excellent weather
conditions, with system temperatures of 115 \,K at 108\,GHz and 210\,K
at 216\,GHz on the $T_{\rm A}^*$ scale.  Dual polarization receivers
were used at both frequencies with the 256 $\times$ 1 MHz filterbanks
on each receiver.  The observations were made with a wobbling secondary
in beam switch mode
with a wobbler throw of 70\arcsec \ in azimuthal
direction and a wobbling frequency of 0.5 Hz.  
Pointing was checked every 60 -- 90 minutes on the nearby
quasars 2251+158, and was very good, with rms offsets of less than
$3$\arcsec \ on average.  At the end of the observations, the
frequency tuning was checked by observing Orion.  The IRAM forward
efficiency, $F_{\rm eff}$, was 0.95 and 0.91 at 115 and 230\,GHz and
the beam efficiency, $B_{\rm eff}$, was 0.75 and 0.54, respectively.
The half-power beam size was 22$^{\prime\prime}$ at 110\,GHz and
11$^{\prime\prime}$ at 215\,GHz. The CO spectra and luminosities are
presented on the main beam temperature scale ($T_{\rm mb}$) which is
defined as $T_{\rm mb} = (F_{\rm eff}/B_{\rm eff})\times T_{\rm A}^*$.
For the data reduction, the spectra were summed and a constant
continuum level was subtracted.

\subsection{Optical observations}

The spectra of the H{\sc ii} regions towards SQ~B were obtained in
July 2003 with the MOSCA instrument installed on the Calar Alto 3.5m
telescope. Using the multi-slit capabilities of the spectrograph, we 
obtained about 70 spectra of intergalactic H{\sc ii} regions in the
field of SQ. Each region was observed with
a medium (R1000, 1.5 \AA/pixel) and a low (B500,
2.9 \AA/pixel) resolution grism.   
Special care was taken in positioning the slits and
extracting the spectra. Most H{\sc ii} regions have an extremely faint
continuum, so that they could not be used to trace the spectrum along
the dispersion direction. Hence, the aperture corrections were done by
manually fitting the positions of the emission lines.  The rest of the
data reduction was carried out in a standard way using the IRAF
software. Details of our spectroscopic study will be presented in a
forthcoming paper (Duc et al., in preparation), 
and here we only report on results relevant to
SQ~B.

\begin{figure*}
\centering
\resizebox{17.cm}{!}{\rotatebox{0}{\includegraphics{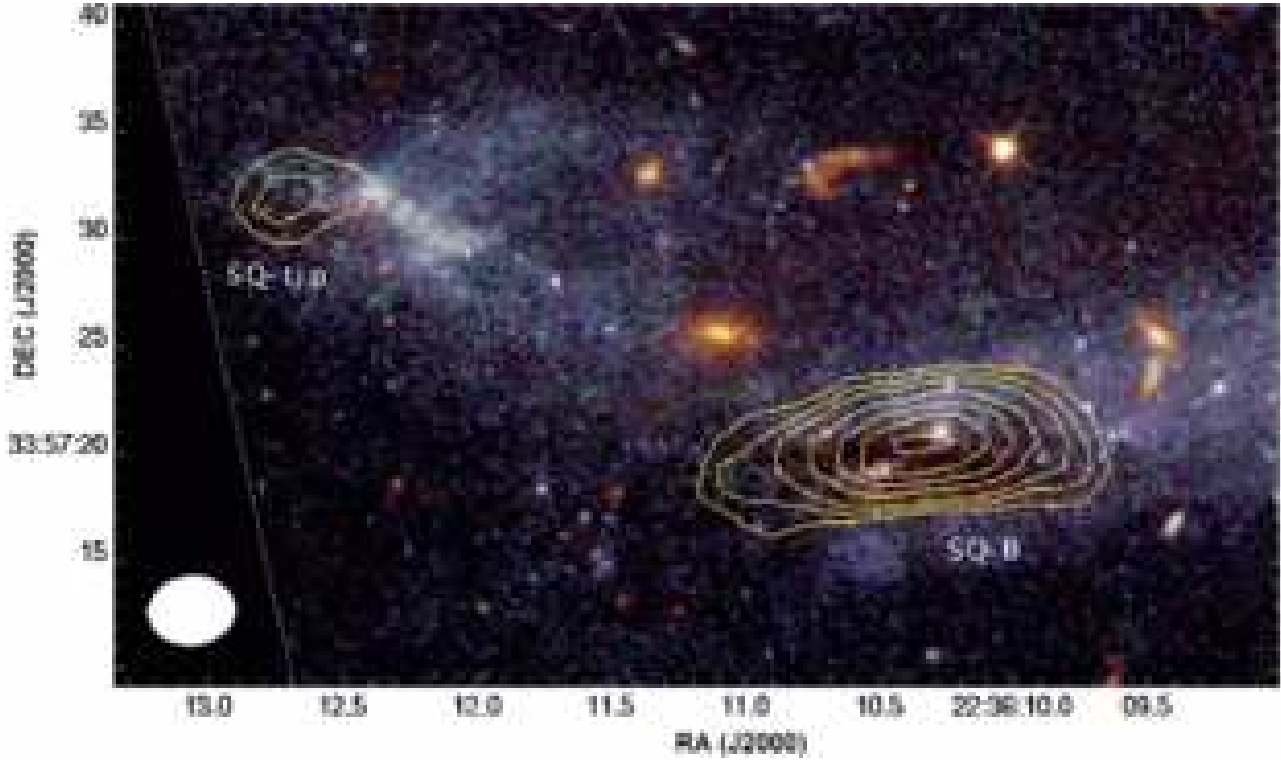}}} 
\caption{
  Contours of the PdBI observations overlaid on a Hubble Space Telescope
  (HST) image  \citep{Gallagher01}.  The velocity
  integration was carried out from 6570 to 6670 \kms.  The contour
  levels presented are at 2$\sigma$, 3$\sigma$, 4$\sigma$, and then increase 
  in intervals of 2$\sigma$.  Note that 2$\sigma$(=0.18 Jy \kms
  beam$^{-1}$) corresponds to a surface density of $2\times 10^{20}$
  cm$^{-2}$, and the maximum contour corresponds to $1.2 \times
  10^{21}$ cm$^{-2}$. 
The beam is shown  in the lower left corner.
}
\label{sq-hst-pdb}
\end{figure*}


\section{Results}

\subsection{Distribution and mass of the molecular gas}

\begin{figure*}
\centering
\resizebox{17.cm}{!}{\rotatebox{270}{\includegraphics{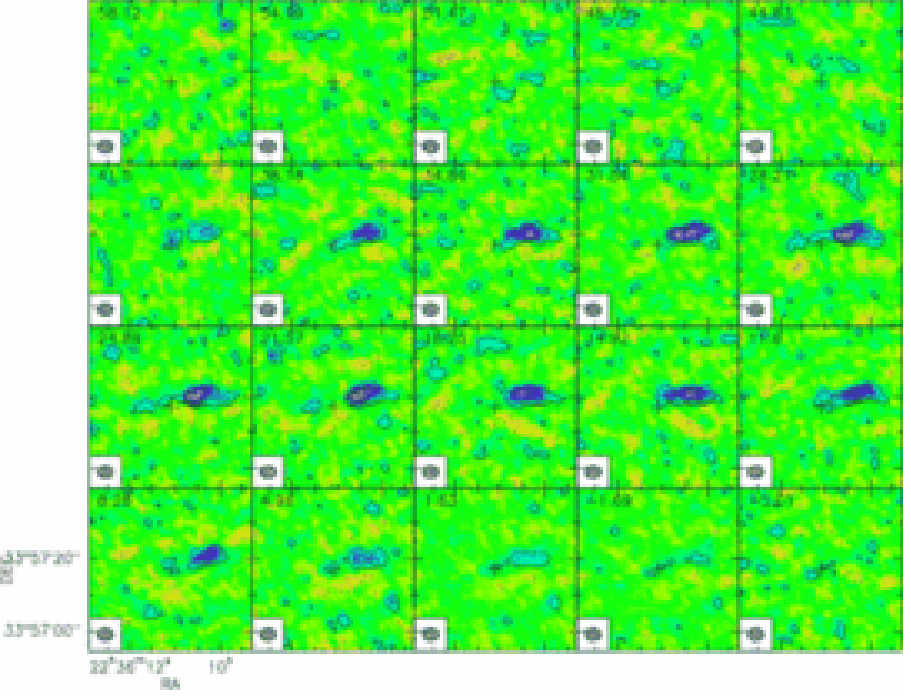}}} 
\caption{Channel maps with a velocity resolution of 3.3 \kms. The
velocity is indicated at the upper left of each panel, given relative
to 6600 \kms. The beam shape is given in the lower left corner. 
The contour levels start at 6.7 mJy/beam (1$\sigma$) and increase in
intervals of 1$\sigma$. The cross indicates the center of the observations at
 $\alpha _{\rm J2000}=22^{\rm h}36^{\rm
  m}11.1^{\rm s}$ and $\delta _{\rm J2000}=33^{\circ }57'17''$.
}
\label{channel-l02}
\end{figure*}

Fig. \ref{sq-hst-pdb} shows the velocity integrated intensity map of
the PdBI observations overlaid over an optical 
Hubble Space Telescope (HST) image \citep{Gallagher01}.  The CO emission
is distributed mainly in two regions. The most prominent emission
region has an elongated shape that roughly follows the tidal tail with
the maximum centered on a dust lane crossing SQ~B. Its overall
extent is $\sim$18\arcsec $\times$7\arcsec or $\sim7\times3$ kpc.  A
much fainter emission region is visible at the tip of the tidal tail,
SQ~tip, slightly offset to the west from the tail and also coinciding
with a dust lane.
The integrated emission at this position is weak, but CO
emission is clearly visible in the individual spectra.  
The molecular gas distribution  with dimensions
of  about 5\arcsec $\times$7\arcsec (2
$\times$ 3 kpc) is only barely resolved by our beam. The channel maps (Fig.
\ref{channel-l02}) reveal that the distribution of the CO within SQ~B
is not completely smooth but shows substructures appearing at
different velocities.

In order to derive the total molecular gas masses in the different
regions we have summed the spectra from the low velocity resolution
data cube over the area where emission is visible in individual
spectra. These summed spectra were then velocity integrated over the
velocity range where emission was visible (6570 to 6670 \kms) and the
molecular mass was calculated assuming (as throughout this paper) the
Galactic conversion factor of N(H$_2$)/I$_{\rm CO}$ = $2 \times
10^{20} {\rm cm}^{-2}$ (K \kms)$^{-1}$.  
The formula used to calculate the molecular gas mass from the
velocity integrated flux, $S_{\rm CO}$, is:

\begin{equation}
M_{\rm H_2} = 7.8 \times 10^3 \left(D/{\rm Mpc}\right)^2 S_{\rm CO} M_\odot,
\end{equation}
where $D$ is the distance in Mpc. This formula corresponds to the one
derived in \citet{Braine01} (eq. 4) multiplied
by a factor 0.73 in order to neglect the helium mass.
The derived line fluxes and molecular gas masses are listed in 
Table \ref{mol_mass}.

The total molecular mass detected by the PdBI observations is $(3.3 \pm
0.7) \times 10^8$ \msun.  The total mass derived from the IRAM 30m
observations \citep{Lisenfeld02}, is $(7.0 \pm 1.4) \times 10^8$
\msun, showing that the PdBI observations pick up only about half of the
total CO emission, the rest being distributed on
scales larger than about 10--15\arcsec \ and therefore resolved out by the
interferometer.

In order to better compare the spatial distribution of the compact and
smoothly distributed emission in SQ~B and SQ~tip, 
we have co-added in each case the
CO emission of the 30m observations over 2 beams, covering the area
where emission was detected by the PdBI in SQ~B and SQ~tip.  Table
\ref{mol_mass} summarizes the molecular gas masses derived from the
IRAM 30m and PdBI in the individual regions.  Due to the poor spatial
resolution of the 30m observations, the corresponding regions do not
coincide completely (the 30m data covers a larger area).  The molecular 
gas masses observed at SQ~B with PdBI is 
$65\pm 22$\% of the mass detected with the 30m telescope in a slightly larger
area.  At the tip of the tidal tail, the PdBI observations
detect a slightly lower fraction of about $40\pm 14$ \% of the total CO.
Thus, within the uncertainties, 
the fraction of molecular gas picked up 
by the interferometer  is not significantly different in both regions.


\begin{table}
\caption{Molecular gas masses in different regions}
\begin{tabular}{cccc}
\hline
Region & \multicolumn{2}{c}{PdBI} & IRAM 30m \\
   &  Flux & $M_{\rm H_2}$ & $M_{\rm H_2}$ \\
     &  [Jy \kms] & $10^8$ \msun  & $10^8$ \msun \\
\hline
SQ~B  &  4.4 $\pm$ 1.0   & 2.5 $\pm$ 0.6  &  3.9   $\pm$ 0.9 \\
SQ~tip & 1.5 $\pm$ 0.4   & 0.8 $\pm$ 0.2  &  1.9  $\pm$ 0.5 \\
Total & 5.9$\pm$ 1.3      & 3.3 $\pm$ 0.7  & 7.0 $\pm$ 1.4$^{(*)}$ \\

\hline
\end{tabular}

The error includes both the formal error based on  the 
rms noise in the integrated spectrum, the 
uncertainty in the calibration (20\%) 
and, for the PdBI data, 
an estimate for uncertainties due to different parameters 
in the data reduction, above all during the cleaning process (10\%).

(*) Refers to the central 15 beams observed with the 30m telescope
\citep[see][]{Lisenfeld02}.
\label{mol_mass}
\end{table}


\subsection{$^{13}$CO in SQ~B}

\begin{figure}
\resizebox{\hsize}{!}{\rotatebox{270}{\includegraphics{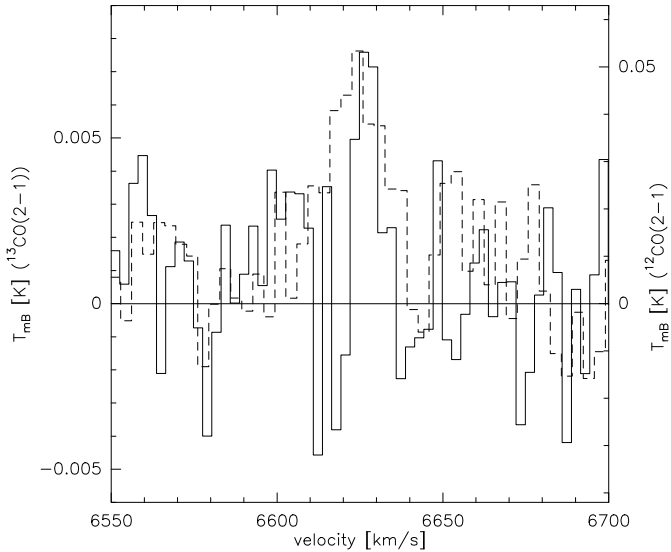}}} 
\caption{The $^{13}$CO(2--1) spectrum at SQ~B (solid line, left temperature
  scale is applicable) compared to the spectrum of the $^{12}$CO(2--1)
  line (dashed line, right temperature scale) at the same
  position.  
The velocity resolution in 2.78 \kms .
}
\label{13co}
\end{figure}

The IRAM 30m observations of $^{13}$CO(2--1) resulted in a 3$\sigma$
detection. As shown in Fig.~\ref{13co} the
central line velocity agrees very well with the
$^{12}$CO(2--1) line observed at the same position in
\citet{Lisenfeld02}, so we are confident that the line
is really detected. The velocity integrated intensity of
$^{13}$CO(2--1) is $0.06 \pm 0.02$ { K \kms}.  However, no $^{13}$CO(1--0) was
detected, down to a rms noise level of 1mK at a velocity resolution of
10 \kms, which corresponds to an upper limit of 0.04 K \kms.

We can
calculate the line ratios between the $^{12}$CO and $^{13}$CO
emission from the velocity integrated intensities of the $^{12}$CO lines at
the same position: I$(^{12}$CO(1--0)) = $1.1 \pm 0.1$ and
I$(^{12}$CO(1--0)) = $1.5 \pm 0.2$ \citep{Lisenfeld02}.
We find that $^{12}$CO(1--0)/$^{13}$CO(1--0) $> 28$ and
$^{12}$CO(2--1)/$^{13}$CO(2--1) $= 25\pm 9$.  These values are much
higher than the value of $\sim$10 that is typically found in
disks of spiral galaxies \citep[e.g.][]{Paglioni01}, and it is close
to the values found for luminous infrared galaxies \citep{Casoli92}.

\subsection{Properties of the ionized gas}

In SQ~B, four individual \hii \ condensations are 
clearly visible on the \halpha \ narrow band
image (see Fig.~\ref{hst-halpha-hi}). We obtained an 
optical spectrum of each one of them. 
All spectra are characterized by a weak continuum and 
strong emission lines. The
equivalent width of the H$\alpha$ line exceeds 200\AA.
Optical velocities (in the Local Standard of Rest, LSR, system) 
were derived from the 
redshifts of the Balmer lines in the medium-resolution
spectra. They range between 6575-6765 \kms. The brightest \hii \  
region has an average
velocity of 6625 \kms \ (range: 6575 - 6655 \kms).

Assuming an intrinsic value of 2.85 for the Balmer decrement
H$\alpha$/H$\beta$, in agreement with Case B recombination models
\citep{Osterbrock89},
we derive an $A_{\rm
  V}$ as high as 3 mag for SQ~B, consistent with the presence of a
dust lane.  
The oxygen abundance of SQ~B was determined
empirically based on the [N{\sc ii}]/H$\alpha$ flux ratio and the
calibration of \citet{vanZee98}. We obtained a value of 12+log(O/H) =
8.7 which is nearly solar \citep[given the latest value of the oxygen
abundance in the Sun,][]{ Allende01}.  This value is
rather high: H{\sc ii} regions along tidal tails have typically a
12+log(O/H) = 8.4--8.6 \citep{Duc04}.  It implies that the gas that made SQ~B was
not expelled from the outer regions of its parent galaxy, but 
rather from inside the disk.

\subsection{Comparison between the ionized and neutral gas} 

\begin{figure}
\resizebox{\hsize}{!}{\rotatebox{0}{\includegraphics{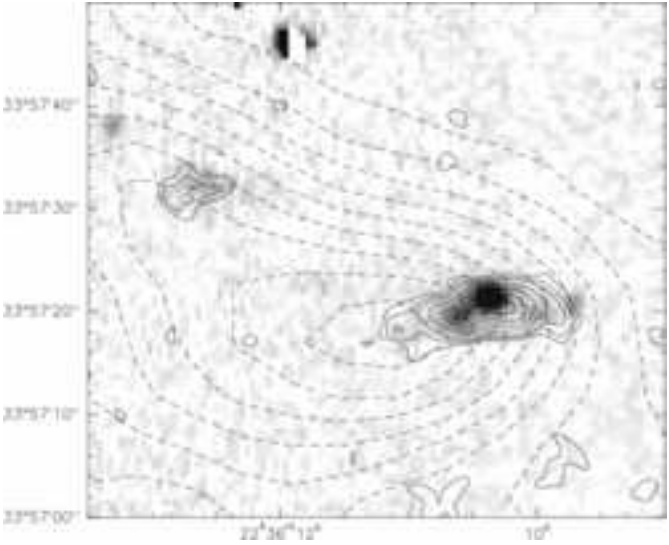}}} 
\caption{
Contours of the PdBI observations overlaid over an
H$\alpha$ image \citep[from][]{Iglesias01}.
The CO contours are as in Fig.~\ref{sq-hst-pdb}.
The dashed contours give the HI emission with
levels starting at 
$1 \times 10^{20}$
atoms \cm, and continuing in intervals of   $1 \times 10^{20}$
until $9 \times 10^{20}$  atoms \cm.
}
\label{hst-halpha-hi}
\end{figure}

In Fig. \ref{hst-halpha-hi} we present a contour overlay of the CO
emission, detected by the PdBI, to the H$\alpha$ emission
\citep[][greyscale]{Iglesias01}. Contours of the HI emission detected
by \citet{Williams02} are also shown with dashed lines.  The exact spatial
coincidence of the CO emission with the H$\alpha$ emission in both SQ
B and SQ tip is striking.  Since dust is most likely coexisting with
the molecular gas, this close correspondence is consistent with the
high extinction obtained from the optical spectroscopy.  The presence
of dust is also clearly visible in SQ B in the optical image, where
the peak of the CO emission coincides with a dust lane (see Fig.
\ref{sq-hst-pdb}).  We can estimate the expected extinction at the
peak of the CO emission from the gas surface density adopting a
Galactic extinction coefficient of $4.5 \times 10^{-22}$ mag cm$^{-2}$
in the R-band \citep{Draine03}.  From the HI data we derive a column
density of 7 $\times 10^{20}$ cm$^{-2}$ for SQ B and our PdBI CO data
correspond to a peak of 17 $\times 10^{20}$ cm$^{-2}$ at the same
region.  Assuming a foreground screen of dust, this implies an
extinction of $A_{\rm V} = 2.3$ mag, slightly lower but in the 
similar range as the value found from the Balmer decrement.
 In SQ~tip the observed column
densities of $N_{\rm HI} = 6 \times 10^{20}$ cm$^{-2}$ and $N_{\rm H_2} = 8
\times 10^{20}$ cm$^{-2}$ result in a somewhat lower extinction of
$A_{\rm V} = 1.3$ mag.

The eastern end of the CO emission in SQ~B coincides with the HI peak
and extends towards the west in a region with a steep gradient in
the HI distribution as already noted in \citet{Lisenfeld02}. This
gradient goes along the optical tidal arm.  Also in SQ tip, the CO
emission lies at the end of this gradient.

\subsection{Gas kinematics}

\begin{figure}
\resizebox{\hsize}{!}{\rotatebox{0}{\includegraphics{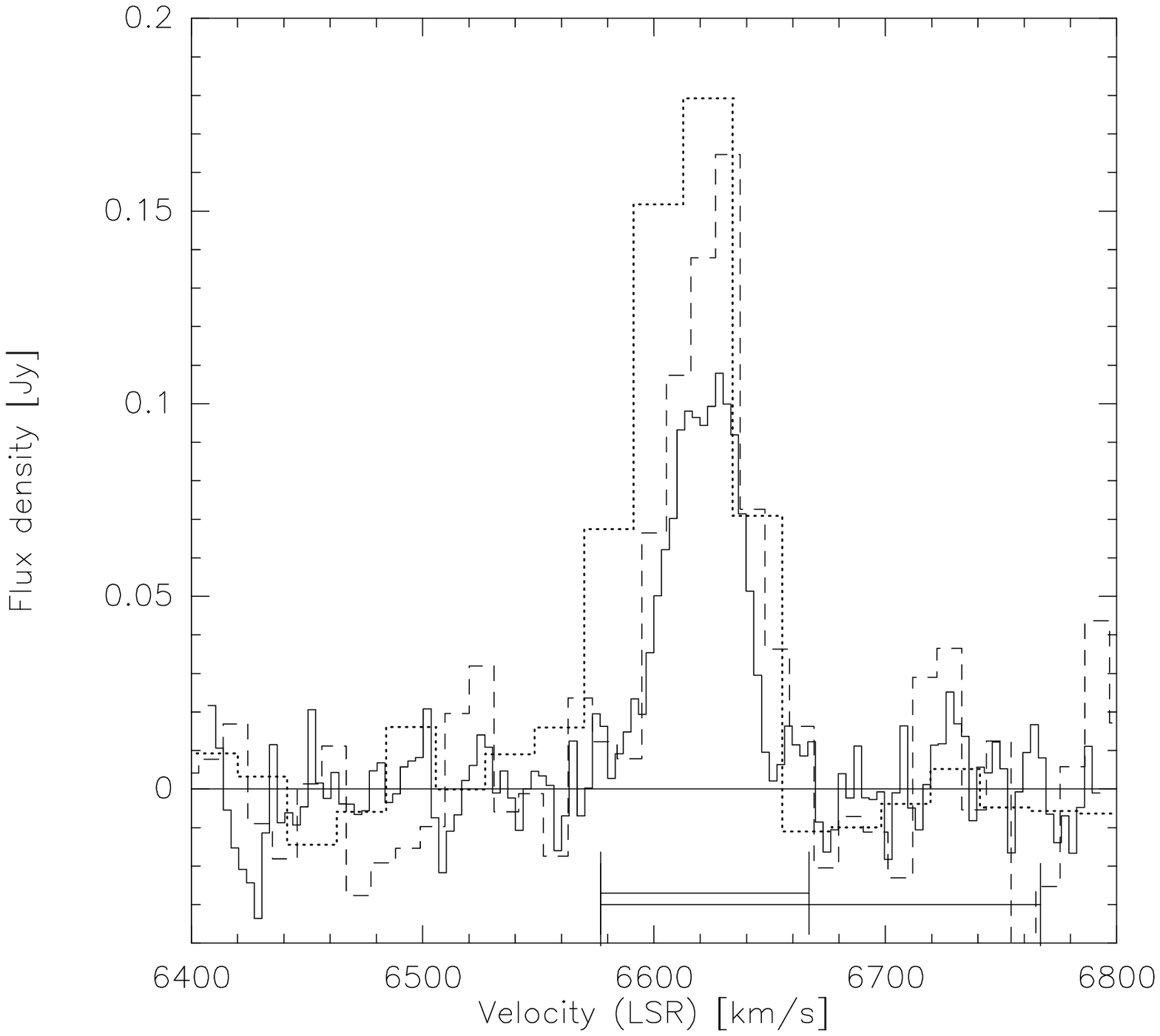}}} 
\caption{
The $^{12}$CO(1-0) spectrum of PdBI 
with a velocity resolution of 3.3 \kms (solid line)
integrated over the emission in SQ~B,
together with the 30m spectrum of this region
 (dashed line) and 
 the HI spectrum (dotted line)
integrated over the same
area as the 30m data.
The HI spectrum is scaled such that for equal flux density with the 30m 
spectrum the corresponding HI surface density is half the one of H$_2$.
The upper horizontal bar at the bottom indicates
the velocity range of the \halpha \ in  the brightest
H{\sc ii} regions, and the lower
horizontal bar gives the velocity range of all \hii \ regions in SQ~B.
}
\label{lines_sqb}
\end{figure}
\begin{figure}
\resizebox{\hsize}{!}{\rotatebox{270}{\includegraphics{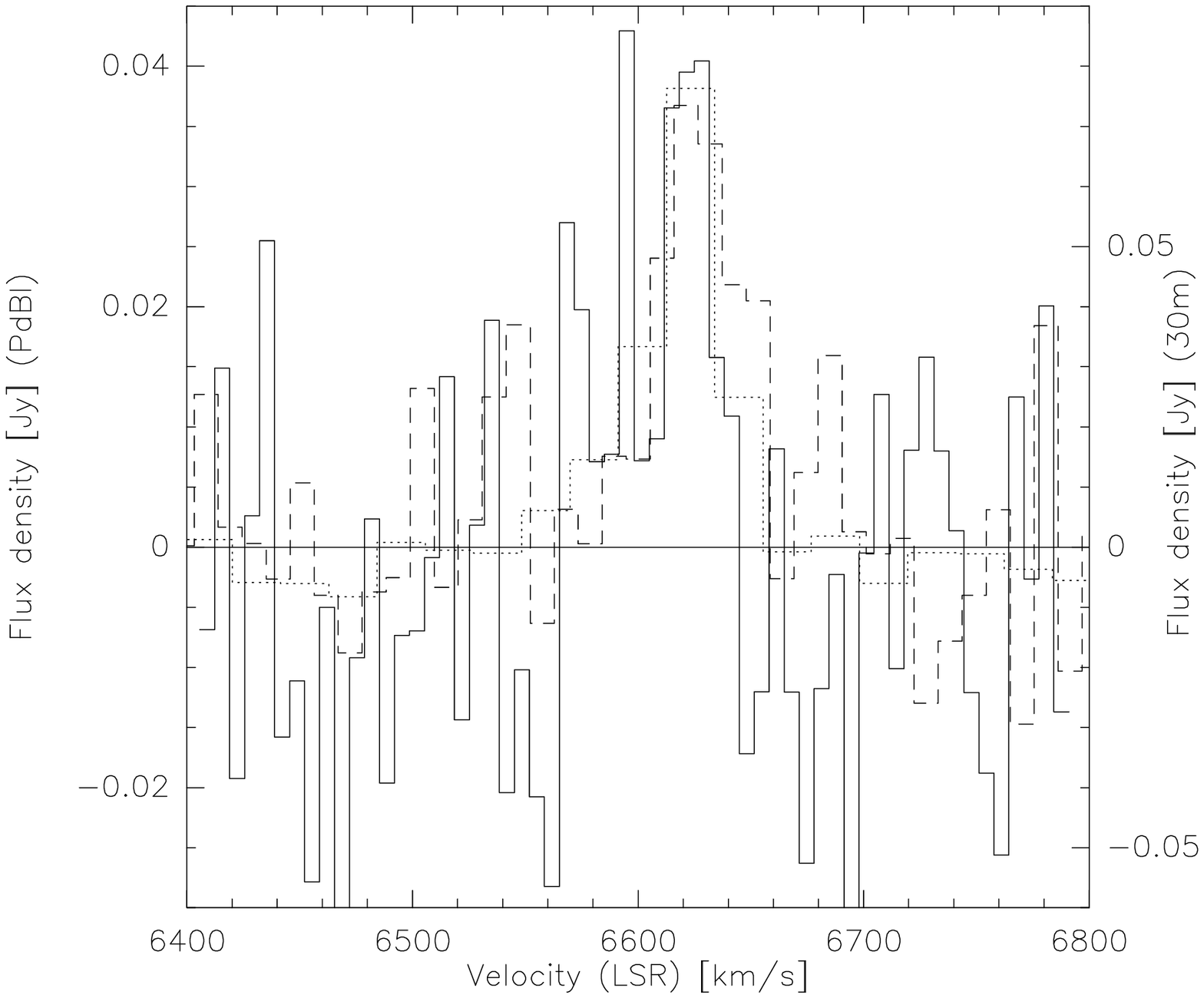}}} 
\caption{
CO spectrum of PdBI  
with a velocity resolution of 3.3 \kms \ (solid line)
integrated over the emission in SQ~tip,
together with the 30m spectrum over this region  (dashed line) and 
 the HI spectrum (dotted line)
integrated over the same
area as the 30m data.
The HI spectrum is scaled such that for equal flux density with the 30m 
spectrum the corresponding HI surface density is 1/7 the one of H$_2$.
}
\label{lines_sqtip}
\end{figure}

In Fig.~\ref{lines_sqb} and \ref{lines_sqtip} we show the spatially
integrated spectra of SQ~B and SQ~tip, observed with the PdBI and the IRAM
30m telescope (solid and dashed lines respectively). In each case the
IRAM 30m spectra are the sum of the two beams that cover the area with
CO emission detected by PdBI.  In addition we show with a dotted line
the total HI spectra, co-added over the same area.

Clearly in SQ~B both CO spectra agree very well in shape, central
velocity and line width. The same is true for the total
emission in this region detected with the 30m telescope \citep[summing
the emission of 15 beams that cover an area of about 60\arcsec
$\times$ 40\arcsec, see][Fig. 6]{Lisenfeld02} which extends over a larger
region.
The HI line has the same central velocity as the CO emission but it is
slightly broader, which could be due to the poorer velocity resolution
of this dataset.  
The central velocity and
velocity range found for the molecular and
atomic gas agree perfectly  with the \halpha \ emission
of the strongest \hii \ region (see Tab.~\ref{vel-gas}). 
The range of the other \hii \ regions
extends to somewhat higher velocities.

Even though the CO spectra of SQ~tip are noisier (see Fig.~\ref{lines_sqtip}),
we also see that the 
central velocity and line shape are in very good agreement
between the CO spectrum from PdB, the CO spectrum from the
30m, and with the HI line. 

\begin{table}
\caption{Kinematics of the molecular and ionized gas}
\begin{tabular}{ccccc}
\hline
 & \multicolumn{2}{c}{CO} & \multicolumn{2}{c}{\halpha} \\
   &  central  &  velocity & central  & velocity  \\
   &  velocity & range & velocity & range \\
   &  [km/s]  &   [km/s]  &   [km/s]  &   [km/s]  \\
\hline
SQ~B & 6625  & 6575--6665 & 6625$^{(1)}$  & 6575--6655$^{(1)}$ \\
     &       &            &  --    & 6575--6765$^{(2)}$  \\   
SQ~tip & 6625 & 6565--6640 &  -- &  -- \\
\hline
\end{tabular}

$^{(1)}$ For the brightest \hii \ region coinciding with 
the peak of CO

$^{(2)}$ Refers to all 4 \hii \ regions 
\label{vel-gas}
\end{table}

\begin{figure}
\centering
\resizebox{\hsize}{!}{\rotatebox{0}{\includegraphics{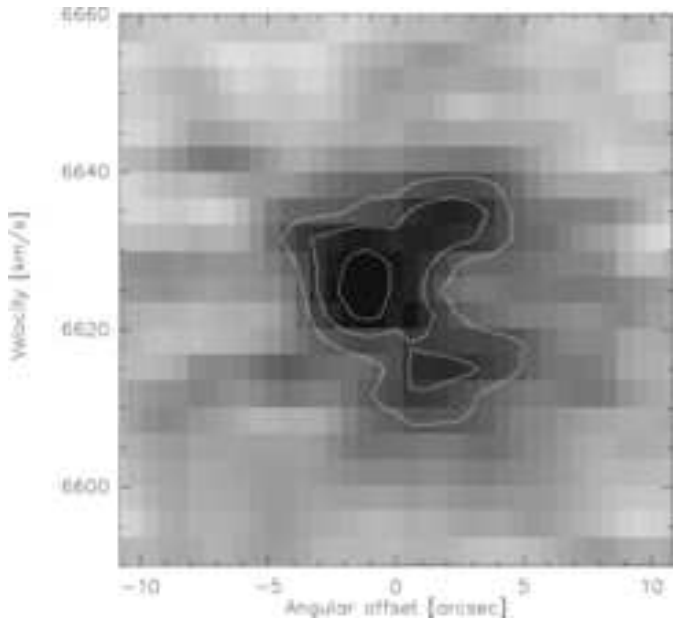}}} 
\caption{
  Position-velocity diagram at SQ~B along the east-west direction with a
  velocity resolution of 3.3 \kms. The contour levels start at 20 mJy/beam 
(3$\sigma$) and increase by 6.7 mJy/beam (1$\sigma$). }
\label{pos-vel}
\end{figure}

We have inspected the CO data cube from PdBI and searched for velocity
gradients by inspecting the channel maps
and by producing various position-velocity (pv) 
diagrams at
various position angles.    
In SQ tip the emission is too weak
to study the kinematics in detail.  
In Fig.~\ref{pos-vel} we show the pv-diagram
of SQ~B obtained in the east-west direction. It shows
an interesting V-shape feature, visible  towards the western side
as a gap in velocities in the range between $\sim$ 6615 and $\sim$
6630 \kms .
The interpretation of this feature is not clear.
A possibility could be two components, one showing
a velocity gradient
and a second superposed component with a velocity offset of 25 \kms.
The \halpha \ data also suggest a complex velocity structure, with
large velocity gradients up to 200 \kms \ which need however to be
confirmed by Fabry-Perot observation in order to exclude that they
are artifacts.

\subsection{Star formation rate and efficiency}

\begin{table}
\caption{Star formation rates derived from different tracers}
\begin{tabular}{lllll}
\hline
Tracer & SFR(SQ~B) &  SFR(SQ~tip)  \\
     &[\msun/yr] & [\msun/yr] \\
\hline
\halpha$^{(1)}$   &   $9.6\times 10^{-2}$ &  $2.9\times 10^{-2}$ \\
\halpha$^{(2)}$  &   0.5  & $7.3 \times 10^{-2}$  \\
\halpha$^{(3)}$  &   0.8 & --  \\
15  $\mu$m     & 0.5  & -- \\
Radio continuum & 0.6 & --\\
\hline
\hline
\end{tabular}

$^{(1)}$ No extinction correction

$^{(2)}$ Extinction correction of $A_{\rm V} =2.3$ (SQ~B) and  
$A_{\rm V} =1.3$ (SQ~tip), based on the gas surface density

$^{(3)}$ Extinction correction of $A_{\rm V} =3$ derived from
Balmer decrement

\label{tab-sfr}
\end{table}

The star formation rate (SFR) at the two regions we detected in CO can
be estimated using a number of tracers.  The \halpha \ 
emission is most commonly used, even though it suffers from
extinction. The total \halpha \ emission in SQ B and SQ~tip is $1.2
\times 10^{40}$ erg s$^{-1}$ and $3.6\times 10^{39}$ erg s$^{-1}$,
respectively. The first value is from \citet{Xu99} whose \halpha \ 
image only covered SQ B but not SQ tip, { and is corrected
for N{\sc II} contamination, assuming [N{\sc II}6538]/\halpha=0.4.  The second one}  is from
\citet{Iglesias01}, after calibrating their deep \halpha \ image using
the SQ B flux derived by \citet{Xu99}. We can convert these \halpha \ 
fluxes to a SFR, using the usual formula SFR $ = 3.0 \times 10^{-8}
(L_{\rm H\alpha}/L_\odot)M_\odot {\rm yr}^{-1}$ of \citet{Kennicutt98},
which assumes an Initial Mass Function (IMF) with a Salpeter slope over the
0.1 to 100 \msun \ range.  In Table~\ref{tab-sfr} we present the SFR
estimates for different extinction corrections.

In SQ~B the SFR can also be measured using two extinction-free
tracers, the thermal dust emission and the radio continuum.  The
luminosity at 15 $\mu$m of SQ~B is $L_{\rm 15\mu m} = 8.2\times
10^{8}$ \lsun \ \citep[][after correcting their value to our
assumed distance]{Xu99}.  Applying the empirical relation of
\citet{Roussel01}, SFR $ = 6.5 \times 10^{-9}L_{\rm 15\mu m}$/\lsun,
we obtained a value of 0.5 \msun yr$^{-1}$.

A final estimate for the SFR can be made from the radio continuum.
Emission at 1.40 GHz was detected from SQ~B at a level of $0.6 \pm 0.2$ mJy
\citep{Xu03} corresponding to a radio luminosity of $5.2\times
10^{20}$ W Hz$^{-1}$. The spectral index between 1.40 GHz and 
4.86 GHz of $0.7\pm 0.4$ found by \citet{Xu03} is typical for
spiral galaxies \citep{Condon92}. We assume therefore that,
as on average found for spiral galaxies, synchrotron radiation
is responsible for 90\% of the emission at 1.4 GHz \citep{Condon92}.
With this assumption, we can apply eq. (18) of \citet{Condon92}
to derive a SFR of 0.6  \msun yr$^{-1}$, assuming a Salpeter IMF.

The SFRs in SQ~B derived from the extinction-free tracers (dust
emission and radio continuum) agree surprisingly well and indicate
levels of SF in SQ~B of about 0.5 \msun yr$^{-1}$.  The same values are
derived from the \halpha \ after applying the extinction correction
derived from the gas surface density, suggesting that this value is
correct.   This SFR is very high for an extragalactic object, much higher
than the values observed in other TDGs,
{ most of them of comparable or larger dynamical mass}
\citep{Braine01}.

Based on the above values of the SFR we can derive the gas consumption
time, defined as the molecular gas mass divided by the SFR (i.e. the
inverse of the star formation efficiency).  In SQ~B, the
(extinction-free) SFR of 0.5 \msun yr$^{-1}$ and the molecular
gas mass of $2.5\times 10^8$ \msun (from the PdBI observations), yield
a gas consumption time of 0.5 Gyr.  In SQ~tip, applying the extinction derived
from the gas column density results to a SFR of 0.073 \msun yr$^{-1}$
and with $M_{\rm H_2} = 8 \times 10^7$ \msun \ the corresponding  gas
consumption time is 1.1 Gyr. This estimate
though is rather uncertain given our limited knowledge of the
extinction of the \halpha \ flux { and of the molecular gas conversion
factor}.  
{  Our measurment lies  within the range of gas consumption times
found for spiral galaxies, the average being  0.6 Gyr (adjusted to our value of the conversion
factor) with a dispersion of a factor 3
\citep{Kennicutt98}.}
This result is
still valid even if, as discussed in Section 4.2, our conversion
factor overestimates the molecular gas content in SQ~B. However, in
this case the gas consumption time would be more similar to the values
of 0.25 Gyr with a dispersion of {  a factor of} 2.5 derived for starburst galaxies by
\citet{Kennicutt98}. It is unlikely that the conversion factor is
{\it higher} than the Galactic value  
 due to the
solar metallicity observed in SQ~B.

\section{Discussion}

\subsection{Association with the optical tidal tail of NGC 7319}

On the basis of our observations  we may be able to answer the question:
is the north-eastern HI cloud \citep[called Arc-N in][]{Williams02}
physically associated with the optical tidal tail or not?
This HI cloud overlaps  at its southern end with the optical tidal tail,
but it has been  unclear whether this is simply  a projection effect.
The enormous size of this cloud makes it distinctly different from  what is
observed  in tidal tails in other interacting systems.  
So far, the only argument in favor of 
a physical relation  is 
the  sharp inner edge of the gas cloud 
coinciding in position and shape with the stellar optical tail
\citep{Sulentic01}. 

Our CO observations have shown that  there is very good spatial and
kinematical agreement between the CO and \halpha \ at both SQ~B and SQ~tip.
Furthermore, the line velocities and shapes of CO and HI match very well.
From these results we conclude that the recent SF traced by the \halpha \ 
is indeed fed by the molecular gas observed in CO and both are
physically related to the atomic gas. Strictly speaking, this does not
prove
that these  components are related to the older stars seen in the
optical tidal arm. In principle, the optical tail could be a background object 
to the atomic, molecular and ionized gas, including the recent
SF that it contains.

The main argument against such a projection effect is the fact that the spatial
coincidence between the  blue stellar knots in the optical tail at 
SQ~B and SQ~tip and the \halpha \ emission
occurs at {\it two positions} in the tidal tail. It is
unlikely that this is due to a coincidence and we conclude
that the \halpha \ emission most likely has  its origin {\em within} 
the optical tail.
Since our observations establish a physical relation between the neutral
 and ionized gas, also 
the HI and molecular gas must therefore be related to the optical tail.

These results  allow us to draw a further conclusion.
The optical tidal tail stems from NGC 7319, which points at  this 
galaxy as its progenitor. Since the gas and the optical tail are related,
their origin is expected to be the same which suggests that
the gas also comes from    NGC~7319.
The  solar metallicity observed in SQ~B furthermore indicates that the
gas comes from relatively inner regions of this galaxy since an enrichment by
recent SF is not sufficient to explain these values
\citep[see][]{Braine01}.

\subsection{Physical conditions of the molecular gas}

The high $^{12}$CO/$^{13}$CO line ratio derived both for the (1--0)
and the (2--1) line is similar to the values
found in 
starburst galaxies \citep{Aalto95}, including their extreme forms of 
infrared  luminous (LIRGs) and
ultraluminous (ULIRGs) galaxies \citep{Solomon97}. 
In starburst and (U)LIRGs the
$^{12}$CO emission is believed to originate not from dense,
self-gravitating molecular clouds, but instead from a moderately dense 
inter-cloud medium
with a relatively low opacity \citep{Aalto95,Solomon97}.
Dense molecular gas also exists  in great abundance  as shown 
by the strength of the line emission of dense gas tracers
such as HCN \citep{Gao04}.  Thus, these observations point to a two
component interstellar medium (ISM): dense clouds where  
molecules as HCN
emit and a diffuse intercloud medium which is, in contrast to
normal galaxies, dense enough for the gas to be molecular.
As a result of this  moderate density, the molecular  gas masses 
derived from the CO emission
using  a Galactic conversion factor are a factor of 3--5 too high
\citep{Braine98, Downes98}.

An alternative explanation for the high $^{12}$CO/$^{13}$CO line ratio
is a different abundance of the
$^{13}$C isotope in starbursting galaxies \citep{Casoli92}.
These authors have proposed two possible processes 
that might be responsible for  the underabundance
of  $^{13}$C: (i) the selective production of $^{12}$C in the young
stars formed during the starburst and (ii) the inflow of low-metallicity gas
with a high $^{12}$C/$^{13}$C ratio from the outer regions into the
central regions where the starburst takes place. 
In SQ~B however, we  know that the gas fueling the SF  has a high
metallicity and does therefore not originate from the outer regions
of the parent galaxy so that process (ii) cannot be at work here. 
The  increase in the $^{12}$C/$^{13}$C abundance  ratio due to process (i)
depends very much on the  adopted parameters \citep[see the arguments in
Sect. 5.2 of][]{Casoli92}.
Adopting a realistic value for the gas  mass fraction
converted into stars (20\%) and assuming that the IMF is 
not biased towards massive stars,
we derive an increase in  $^{12}$C/$^{13}$C of not more than 10\%. 
Only in the case of an IMF strongly biased towards massive stars could
this  process
account for the observed high $^{12}$CO/$^{13}$CO line ratio in SQ~B.

In the case of SQ~B it seems therefore that 
a low density/opacity of the ISM  is the most likely explanation for the
low $^{13}$CO/$^{12}$CO line ratio. 
As a consequence it is possible that the
Galactic conversion overestimates the molecular gas masses.
A comparison between SQ~B and starburst  galaxies
is not unreasonable, since the SF activity in SQ~B is indeed high, as
indicated by the high 
 equivalent width of the \halpha \ line of $> 200$\AA.  
More line observations, in particular high-density tracers as HCN, will
be necessary to better understand the properties of the
molecular gas in SQ~B.

%

Interestingly, a high $^{12}$CO/$^{13}$CO has also been found in
a very  different type  of object, in the bridge of the
interacting system Taffy galaxies \citep{Braine03}. Here, the
CO/HCN ratio is very low, in contrast to starburst and (U)LIRGs.  At
the same time only very little SF is found.



\section{Summary and conclusions}

We have presented our analysis of new high resolution ($4.3''\times
3.5''$) CO observations of the eastern tidal tail in SQ obtained with
the interferometer at Plateau de Bure (PdBI), 
as well as optical spectroscopy of the two star forming (SF) regions in the
zone, SQ~B and SQ~tip, along with observations of the $^{13}$CO line
with the IRAM 30m telescope. The main results derived from our
observations are:

(i) We have identified two main CO emitting regions, associated with
regions of SF in the optical tail at SQ B and at the tip of the tidal
tail, SQ tip.
Our PdBI maps trace about half of the total CO emission derived
from single dish 30m IRAM observations \citep{Lisenfeld02}. This
indicates that about half of the molecular gas is distributed on
scales larger than 10--15\arcsec \ for which PdBI is not sensitive.
 The spatial
extent of this latter component is very large, covering an area of
about $16 \times 25$ kpc \citep[see][]{Lisenfeld02}.

(ii) In both regions the CO coincides exactly with \halpha \ emission.
In SQ~B there is furthermore perfect 
 kinematical agreement with the \halpha \ line.
In addition, the line shapes and velocities of CO
and of HI spectra \citep{Williams02} are identical.  
These agreements suggest that the recent SF and molecular and 
neutral gas are related.
\halpha \ and CO emission coincide  with the blue star clusters
 visible in the optical tidal tail. Since this coincidence
occurs at two different positions, it seems unlikely that it is 
a projection and we conclude that a physical relation to the
optical tail exists. This relation suggests that the origin of the
gas is NGC~7319, the galaxy from where the optical tail stems.

(iii) We tentatively detected $^{13}$CO(2--1) in SQ~B
and obtained an upper limit for
$^{13}$CO(1--0).  The $^{13}$CO observations yield high values for
the $^{12}$CO/$^{13}$CO ratio of $\ge 25$, both for the 1--0 and for the 2--1
line, higher than those found in disks of spiral galaxies and comparable
to those found in starburst galaxies.
A low opacity of the $^{12}$CO emitting gas is the most plausible
explanation for this result, however
more molecular line observations are necessary in order to draw
firm conclusions.

(iv) Optical spectroscopy in SQ~B and SQ~tip revealed several knots of
\halpha \ emission. Unlike other TDGs, the metallicity of SQ~B is
roughly solar and the extinction derived from the optical spectra is
high (A$_V$ = 3.0).  
This indicates that the gas at SQ~B has its origin in relatively
inner regions of the disk of the parent galaxy.


(v) From the extinction-corrected H$\alpha$ emission and from extinction
free SF tracers (dust emission and radio continuum) we derive a SFR
of SQ~B as high as 0.5 \msun/yr. 
The resulting gas
consumption time of 0.5 Gyr is in the range found for spiral and starburst
galaxies by \citet{Kennicutt98}.

\begin{acknowledgement}
  
  We would like to thank J. Iglesias-P\'aramo, J. V\'ilchez and C.K. Xu
  for making available their \halpha \ data to us, L. Verdes-Montenegro
  for the HI data, J. Sulentic for helpful discussions, and the referee,
  Alessandro Boselli, for useful comments on the manuscript.
  UL and SL are supported by the Spanish MCyT Grant AYA 2002-03338 and
  by the Junta de Andaluc\'\i a.  VC would like to acknowledge the
  partial support of NASA through contract number 1257184 issued by
  JPL/Caltech and EB acknowledges financial support from CONACyT via
  project 27607-E.

\end{acknowledgement}

{}

\end{document}